# Tunable nonlinear and active THz devices based on hybrid graphene metasurfaces

Tianjing Guo, Christos Argyropoulos*

Department of Electrical and Computer Engineering, University of Nebraska-Lincoln, Lincoln, Nebraska 68588, USA

## ABSTRACT

Graphene is a two-dimensional layer of carbon atoms arranged in a honeycomb lattice, whose outstanding properties makes it an excellent material for future electronic and photonic terahertz (THz) devices. In this work, we design hybrid graphene metasurfaces by using a monolayer graphene placed over a metallic grating, operating in the THz frequency range. Perfect absorption can be achieved at the resonance, where the electric field is greatly enhanced due to the coupling between the graphene and the grating plasmonic responses. The enhancement of the electric field along the graphene monolayer, as well as the large nonlinear conductivity of graphene, can dramatically boost the nonlinear response of the proposed THz device. In addition, the presented enhanced nonlinear effects can be significantly tuned by varying the doping level of graphene. The proposed structure can be used in the design of THz-frequency generators and all-optical processors.

*christos.argyropoulos@unl.edu

## 1. INTRODUCTION

As a two-dimensional (2D) materials with one atom thickness, graphene has been demonstrated to be an ideal platform to achieve plasmonic behavior at lower far-infrared and terahertz (THz) frequencies [1], due to its many exciting electrical, optical, mechanical, and thermal properties [2,3]. Graphene can support surface plasmons at THz frequencies, similar to bulk metals at optical frequencies. However, what is more appealing is that the surface plasmons formed at the surface of graphene are highly confined and tunable, which can lead to exciting applications, such as THz switches and phase shifters [4,5]. Graphene can be patterned to circular or rectangular patches [6,7] and used by itself as the building element of metasurfaces. Moreover, graphene metasurfaces can be created by patterning a graphene monolayer placed on top of an insulator [8], which exhibit extremely small wavelengths and high field confinement along the graphene sheet because of its 2D nature. This favorable feature, compared to dielectric metasurfaces, led to enormous interest in the investigation of graphene metasurfaces.

Graphene's third order THz nonlinearity has been demonstrated theoretically and experimentally to be remarkably strong [9]. The strong third-order nonlinear response originates from the intraband electron transitions [10], as well as the resonant nature of the light-graphene interactions. In recent experiments [11], the Kerr nonlinear susceptibility $\chi^{(3)}$ of graphene was reported to reach a very high value $1.4 \times 10^{-15} m^2/V^2$. Among all the nonlinear process, third-harmonic generation (THG) is most common and investigated widely, where an incident wave $\omega$ interacts with the system to produce a wave with 3 times the incident-wave frequency $3\omega$ [12]. The THG process can be completed with only a single-wavelength source, which is different compared to other nonlinear processes. It has been reported to be produced by graphene and few-layer graphite films but with relative low efficiency despite the large nonlinear graphene susceptibility [13,14]. Graphene metasurfaces can be employed to boost the nonlinear effects, even at THz frequencies. Large field enhancement and confinement can be obtained along the graphene sheet [15], which can dramatically amplify the inherently nonlinear response of the structure. THG efficiency can be greatly enhanced due to the large localized field enhancement, as well as the high nonlinear response of graphene.

In this work, we propose hybrid graphene metasurfaces with a monolayer graphene placed over a metallic grating. Strong coupling and interference between the surface plasmons of graphene and localized plasmons of the metallic grating can be obtained, which can lead to the perfect absorption during the investigation of the linear response of this structure. We further investigate the power flow and field distribution in order to verify this perfect absorption. The THG process can be greatly boosted due to the large field enhancement along with the high inherent nonlinear response of graphene. The THG efficiency can reach to very high values with relatively low input intensity. The most appealing feature is that this nonlinear effect can be tuned by dynamically changing the doping level of monolayer graphene. The proposed graphene metasurface has potentials to be used in the design of THz-frequency generators and all-optical processors.

## 2. DESIGN AND NUMERICAL SIMULATION MODEL

### 2.1 Design of graphene metasurface

The geometry of the proposed graphene metasurface is illustrated in Fig. 1(b). The grating is periodic in the x direction with period $p$ and is assumed to be extended to infinity in the $y$ direction. The height and trench width of the grating are equal to $d$ and $b$, respectively. The ground plane is thick enough to be considered opaque to the impinging THz radiation leading to zero transmission. The grating is made of gold, with THz optical constants calculated by using the Drude model: $\varepsilon_{L,Au} = \varepsilon_\infty - f_p^2 / f(f - i\gamma)$, where $f_p = 2069 THz$, $\gamma = 17.65 THz$ and $\varepsilon_\infty = 1.53$ are derived from fitting the experimental values [16]. Gold can be assumed to exhibit a PEC-like response at low THz frequencies, since it has very high conductivity and the fields minimally penetrate its bulk volume [17,18]. We will include its nonlinear susceptibility $\chi_{Au}^{(3)} = 2.45 \times 10^{-19} m^2/V^2$ at the infrared frequency region [19], in addition to its linear Drude model, when we investigate the nonlinear response of the proposed structure.

The grating is covered by a graphene monolayer sheet, which is a 2D conductive material that can be modeled as a surface current, neglecting its sub-nanometer thickness. The graphene linear conductivity under the random-phase approximation can be derived by the Kubo formula consisting both of interband and intraband electron transitions [20]. We design the proposed device to operate at low THz frequencies below the interband transition threshold $\hbar\omega < 2|E_F|$ [21], where $E_F$ is the Fermi level or doping level of graphene. In this case, the interband transitions are blocked due to the Pauli principle, so only the intraband transitions exist. The linear conductivity arising from the intraband transitions can be calculated by the Drude formalism: $\sigma_g = (e^2 E_F / \pi \hbar^2)[j/(j\tau^{-1} - \omega)]$ [1], where $e$ is the electron charge, $\hbar$ is the reduced Planck's constant, $\omega = 2\pi f$ is the angular frequency, $j$ is the unit imaginary number, and $\tau$ is the relaxation time [22,23], which is assumed to be equal to $\tau = 10^{-13} s$ throughout this work. The proposed structure is illuminated by a TM-polarized wave (the magnetic field in the y direction) with an incident angle θ with respect to the z direction, as shown in Fig. 1 (b). Existing well-established fabrication methods can help to realize the proposed structure. Graphene and metallic grating can be separately fabricated, by chemical vapor deposition on a copper foil and electron-beam lithography, respectively, and then deposit the graphene monolayer on the microscale metallic grating [24,25].

The third-order nonlinear surface conductivity of graphene at THz frequencies is calculated by the formula [26]:

$$\sigma^{(3)} = \frac{i\sigma_0(\hbar v_F e)^2}{48\pi(\hbar\omega)^4} T(\frac{\hbar\omega}{2E_F}), \qquad (1)$$

where $\sigma_0 = e^2/4\hbar$, $v_F = 1 \times 10^6 m/s$, $T(x) = 17G(x) - 64G(2x) + 45G(3x)$, with $G(x) = \ln|(1+x)/(1-x)| + i\pi\theta(|x|-1)$, and $\theta(z)$ is the Heaviside step function. The equivalent third-order susceptibility of graphene can be calculated by using $\chi^{(3)} = -j\sigma^{(3)}/(\omega\varepsilon_0 d)$, where $d \approx 0.34 nm$ is the thickness of monolayer graphene. The values of this nonlinear susceptibility are several orders of magnitude larger than conventional dielectric and metallic bulky nonlinear media. Strong nonlinear graphene response can be expected at lower frequencies and under low Fermi level, as implied in Eq. (1). It has been reported that the equivalent nonlinear susceptibility of graphene can reach to $1.3 \times 10^{-10} m^2 V^{-2}$ at 3 THz, which is much larger than the nonlinear susceptibility $2.45 \times 10^{-19} m^2 V^{-2}$ of gold at near-IR [27]. Therefore, the nonlinear response of the metasurface will be dominated only by the ultrastrong Kerr nonlinear susceptibility of graphene. The high nonlinear properties of graphene makes it a very appealing nonlinear material in THz frequencies.

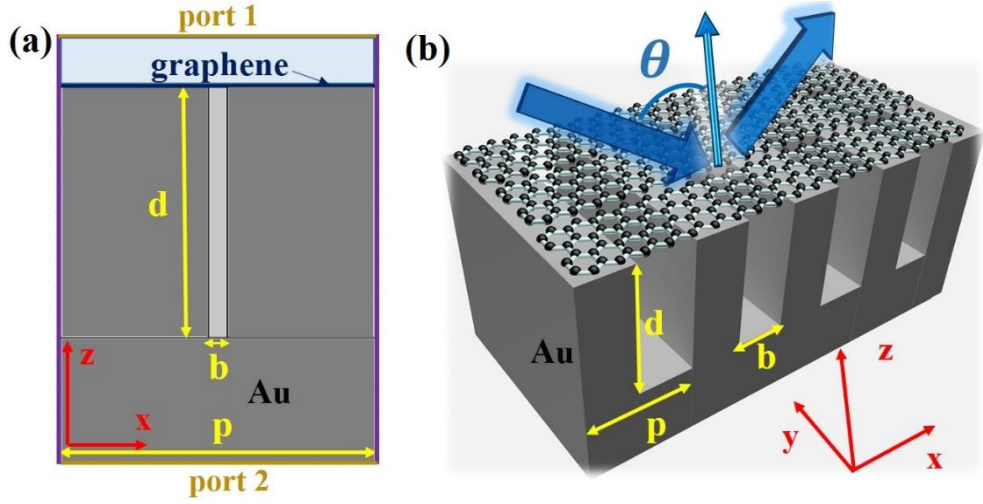

Figure 1. (a) The simplified simulation domain in COMSOL. (b) Schematic of the proposed graphene metasurface. The used PBC and port boundaries are shown with purple and yellow lines in (a), respectively. Graphene is represented with a blue line. Gray areas represent gold.

## 2.2 Numerical simulation model

We employed the full-wave simulation software COMSOL Multiphysics to solve the linear and nonlinear Maxwell's equations and investigate the proposed hybrid graphene metasurface. The simplified simulation model is presented in Fig. 1 (a). The proposed structure is composed of a gold grating covered by a graphene sheet, and graphene exhibits isotropic properties only along its surface. Thus, we can model this structure as a 2D periodic system, where it is assumed to be infinite (or very large) along the y-direction. Periodic boundary conditions (PBC) are employed in both sides at the x-direction and port boundaries are placed up and down in the z-direction to create the incident plane wave. The PBC boundaries are indicated with purple lines and the used ports 1 and 2 are shown by yellow lines in Fig. 1(a).

Graphene is modeled as a surface current due to its planar ultrathin 2D nature, which is represented by a blue line in Fig. 1(a). The graphene's surface current is described by $J = \sigma_g E_{FF}$ and $J = \sigma_g E_{TH} + \sigma^{(3)} E_{FF}^3$ when it is used in the linear and nonlinear simulations [28], respectively. $E_{FF}$ and $E_{TH}$ are the electric fields induced at the fundamental frequency (FF) and the third harmonic (TH) frequency, respectively. Note that these fields are monitored along the entire surface of graphene. The Kerr third-order nonlinear permittivity of gold is given by: $\varepsilon_{NL,Au} = \varepsilon_{L,Au} + \chi_{Au}^{(3)} E_{FF}^2$, where $E_{FF}$ in this case is the induced field at the fundamental frequency monitored along the entire bulk gold volume (composed by both x- and y-components).

To perform the linear simulations at the FF, we employ one electromagnetic wave solver to calculate the reflection and transmission coefficients. The S-parameter calculations were used in COMSOL to compute the total reflected and transmitted power flow through the input (port 1) and output (port 2) ports of the proposed periodic structure. The power delivered to each port (either by reflection or transmission) is calculated by $P = \int_C \vec{S} \cdot \vec{n} = \frac{1}{2} \int_C \text{Re}\{\vec{E} \times \vec{H}^*\} \cdot \vec{n}$, where $\vec{S}$ is the time-averaged Poynting vector, C is the boundary curve of each port, $\vec{n}$ is the boundary norm vector, $\vec{E}$ is the electric field vector, and $\vec{H}$ is the magnetic field vector along the ports. The reflection coefficient of the proposed structure is obtained with $\Gamma = S_{11}$ and the transmission coefficient is $T = S_{21}$. The total absorptance can be calculated by the formula $A = 1 - |\Gamma|^2 - |T|^2$. In the current grating configuration, the transmittance is equal to zero due to the thick gold substrate. Thus, the absorptance is computed by the simpler relationship $A = 1 - |\Gamma|^2$ [29–31].

In order to compute the THG nonlinear radiation, an additional electromagnetic wave solver needs to be included and coupled to the FF solver. The harmonic wave coupling frequency domain method is used in our calculations by nonlinearly connecting several physics interfaces that model the structure in each frequency of the proposed nonlinear process. In the case of THG simulations, COMSOL will solve the nonlinear Maxwell's equations at the TH frequency $f_{TH} = 3f$, where $f$ is the fundamental frequency. In this case, a wave with frequency $3f$ will be generated, when an incident wave with FF $f$ excites the proposed nonlinear graphene-covered grating. The radiated output power of the TH wave is computed by calculating the integral $\int_C \vec{S} \cdot \vec{n}$ over the surface of the structure, where $\vec{S}$ is the Poynting vector crossing the boundary surface C and $\vec{n}$ is the boundary norm vector. Here, the boundary surface C is the top port boundary which is shown by the yellow line in Fig. 1(a). The grating does not reflect or leak any additional higher-order diffraction modes or surface waves and the generated reflected nonlinear radiation is passing only from the top port boundary, which is placed in the far field.

## 3. RESULTS AND DISCUSSION

### 3.1 Perfect absorption and power flow

We first investigate the linear response of the proposed hybrid graphene metasurface. The absorptance versus the frequency is plotted in Fig. 2(a) at normal incident illumination. The grating parameters used here are *p=8um, b=0.6u, and d=8um*. The Fermi level of the graphene sheet on top of the grating is 0.12 eV. From Fig. 2 (a), we find one pronounced perfect-absorption peak at the resonance of the proposed graphene metasurface. At this resonant point (8.95 THz), a magnetic plasmon mode is formed due to the generation of highly localized magnetic fields inside the grating's trench accompanied by high electric fields that are expected to boost nonlinearities. We further calculate the absorptance of a plain metallic grating. The result is shown by red line in Fig. 2 (a). It is interesting that a substantial frequency blueshift is obtained when graphene is introduced over the grating. The addition of graphene will also lead to dynamic tunability of the absorption resonant frequency. Figure 2 (b) demonstrates the electric-field-enhancement distribution along the structure at the resonant frequency, by calculating the ratio $|E/E_0|$, where $E_0$ is the amplitude of the incident electric field. The electric field can be enhanced by approximately 18 times inside the grating's trench. The perfect absorption obtained indicates a strong coupling and interference between the terahertz plasmons of graphene and the metallic grating.

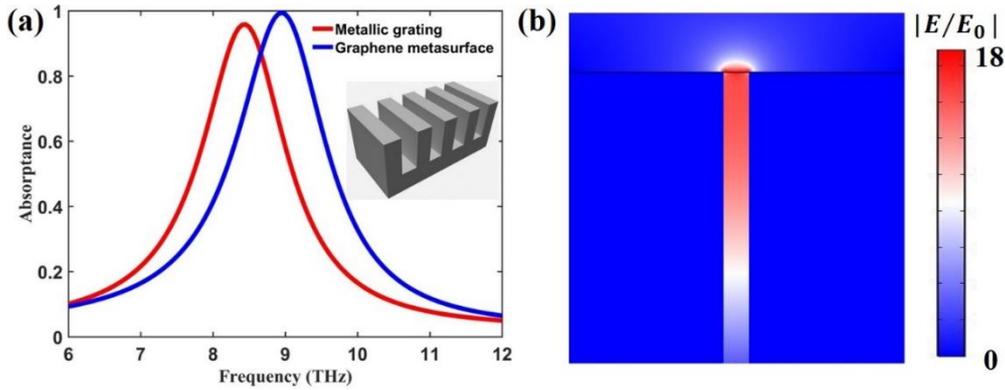

Figure 2. (a) The computed absorption spectra of the proposed metasurface. The absorption spectrum of the metallic grating without graphene on top is plotted as a red line. The results are obtained for the grating parameters *p=8um, b=0.6u, d=8um*. The Fermi level of monolayer graphene is 0.12 eV. The inset is the schematic of the metallic grating. (b) The computed electric-field-enhancement distribution at the resonant frequency of the graphene metasurface.

We calculate the power flow versus frequency at the interface between air and the grating's gold material, in order to demonstrate the perfect absorption obtained by the proposed graphene metasurface, by integrating the time-averaged Poynting vector given by: $P = \int_C \vec{S} \cdot \vec{n} = \frac{1}{2} \int_C \mathrm{Re}\{\vec{E} \times \vec{H}^*\} \cdot \vec{n}$, where $\vec{S}$ is the time-averaged Poynting vector, C is the boundary curve at the gold/air interface extended along the gold part of the grating and not along the air upper part of the corrugations, $\vec{n}$ is the boundary norm vector along this interface, $\vec{E}$ is the electric field vector and $\vec{H}$ is the magnetic field vector. The skin depth of gold around the presented perfect absorption resonance frequency (8.95 THz) is approximately $\delta = 0.02um$ [17] and no power is expected to penetrate inside gold below this ultrathin skin depth. Figure 3 (a) presents the power going into gold versus frequency by integrating the time-averaged Poynting vector over three boundary curves along different depths inside the gold part of the grating. The red line represents exactly the interface between air and gold, the green line is placed at approximately the skin depth of gold (0.02um), and the blue line is located deep inside gold and five times larger compared to the gold's skin depth (0.1um). It is found from Fig. 3 (a) that the power flow is much smaller and almost zero deep inside gold (blue line) than the power at the interface (red line). In addition, the minimum of the power flow at the absorption resonance frequency in Fig. 3 (a) proves that most power goes into the grating's trenches and not inside gold at this frequency point, as it is shown in Fig. 3 (b). The power can fully penetrate and interact with the ultrathin 2D graphene placed on top of the gold grating corrugations, and this can be further derived by observing Fig. 3 (c), where the arrows represent the power flow at the resonance. It can be clearly seen in Fig. 3(d) that all the incident power is fully absorbed inside the elongated trenches of the metallic grating, since the power flow arrows become smaller as they travel deeper inside them. The majority of the field and power enhancement happens exactly at the interface between the grating trenches and air, where graphene is located. Hence, the power of the incident radiation, which is fully absorbed into the corrugations' trenches, is strongly interacting with graphene that leads to the enhanced nonlinear response.

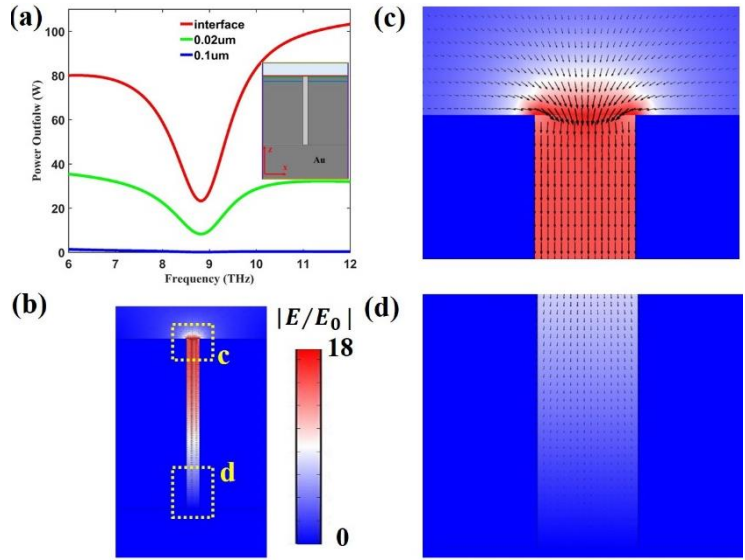

Figure 3. (a) Computed power flow at the interface between air and gold (red line), almost equal to the gold's skin depth (green line), and four times larger (blue line) than the gold's skin depth. The inset shows a schematic with the corresponding positions where the power is calculated. (b) The computed absolute value of the electric field enhancement distribution and the power flow arrows inside the structure at the resonance frequency. The power flow is dissipated and fully absorbed inside the air corrugations' trenches. The results are obtained for grating parameters *p=8um, b=0.6um,* and *d=8um*. The graphene's Fermi level is equal to 0.12 eV.

### 3.2 Tunable nonlinear effects

We next investigate the nonlinear response of the proposed structure by computing the THG conversion efficiency (CE), which is a more-quantitative way to describe the THG power strength. CE is defined as the ratio of the radiated THG power outflow $P_{out,TH}$ to the input FF power $P_{in,FF}$: $CE = P_{out,TH}/P_{in,FF}$ [32]. Equation (1) implies that stronger nonlinear response can be obtained with less-doped graphene. The THG CE will also be affected by the FF since the nonlinear surface conductivity of graphene given by Eq. (1) is inversely proportional to the radial frequency. Figure 4 confirms our conclusion. We calculate the THG CE of the proposed nonlinear structure by sweeping the Fermi level of graphene and the fundamental frequency. The geometric parameters are fixed to *p=8um, b=0.6u,* and *d=8um*. The input intensity of the FF waves is fixed to the low value of $20kW/cm^2$. Clearly, the THG CE decreases as the Fermi level is increased or in the case of off-resonance operation. The maximum THG CE is obtained for $f_{FF} \approx 8.8 THz$ and slightly doped graphene with $E_F = 0.1$ eV. Note that low-doped graphene, which is easier to produce, can lead to enhanced nonlinear effects with the proposed configuration. It is possible to tune the THG nonlinear waves by dynamically changing the Fermi level of graphene sheet. In addition, the computed THG CE can reach very high values by exciting the proposed hybrid graphene metasurface with low input intensities [33].

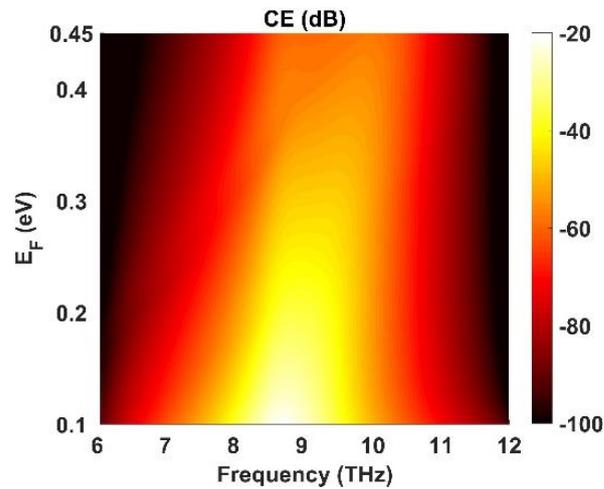

Figure 4. The computed THG CE on a logarithmic scale at normal incidence as a function of the fundamental frequency and graphene's Fermi level of the proposed graphene metasurface.

## 4. CONCLUSIONS

To conclude, this work has presented tunable nonlinear effects with a hybrid graphene metasurface, which is formed by a monolayer graphene placed over a metallic grating. Perfect absorption can be obtained with the proposed design due to the strong coupling between the THz plasmons of graphene and the metallic grating. Large nonlinear conductivity of graphene plays an important role in the great enhancement of the THG process. In addition, the nonlinear response can be dynamically tuned just by varying the doping level of graphene. Note that the proposed structure can be realized with commonly used fabrication techniques and can be used in the design of terahertz-frequency generators and all-optical processors. During the conference, active graphene metasurfaces will also be presented where graphene depicts gain response [34].

## REFERENCES

1. M. Jablan, H. Buljan, and M. Soljačić, "Plasmonics in graphene at infrared frequencies," Phys. Rev. B **80**(24), 245435 (2009).
2. K. S. Novoselov, V. I. Fal′ko, L. Colombo, P. R. Gellert, M. G. Schwab, and K. Kim, "A roadmap for graphene," Nature **490**(7419), 192–200 (2012).
3. Q. Bao and K. P. Loh, "Graphene Photonics, Plasmonics, and Broadband Optoelectronic Devices," ACS Nano **6**(5), 3677–3694 (2012).
4. P.-Y. Chen, C. Argyropoulos, and A. Alu, "Terahertz Antenna Phase Shifters Using Integrally-Gated Graphene Transmission-Lines," IEEE Trans. Antennas Propag. **61**(4), 1528–1537 (2013).
5. P.-Y. Chen, C. Argyropoulos, M. Farhat, and J. S. Gomez-Diaz, "Flatland plasmonics and nanophotonics based on graphene and beyond," Nanophotonics **6**(6), 1239–1262 (2017).
6. Q. Q. Zhuo, Q. Wang, Y. P. Zhang, D. Zhang, Q. L. Li, C. H. Gao, Y. Q. Sun, L. Ding, Q. J. Sun, S. D. Wang, J. Zhong, X. H. Sun, and S. T. Lee, "Transfer-free synthesis of doped and patterned graphene films," ACS Nano (2015).
7. T. Guo and C. Argyropoulos, "Broadband polarizers based on graphene metasurfaces," Opt. Lett. **41**(23), 5592 (2016).
8. L. Ju, B. Geng, J. Horng, C. Girit, M. Martin, Z. Hao, H. a Bechtel, X. Liang, A. Zettl, Y. R. Shen, and F. Wang, "Graphene plasmonics for tunable terahertz metamaterials," Nat. Nanotechnol. **6**(10), 630–634 (2011).
9. K. Yang, S. Arezoomandan, and B. Sensale-Rodriguez, "The linear and non-linear THz properties of graphene," Terahertz Sci. Technol. **6**(4), 223–233 (2013).
10. M. M. Glazov and S. D. Ganichev, "High frequency electric field induced nonlinear effects in graphene," Phys. Rep. **535**(3), 101–138 (2014).
11. E. Hendry, P. J. Hale, J. Moger, A. K. Savchenko, and S. A. Mikhailov, "Coherent Nonlinear Optical Response of Graphene," Phys. Rev. Lett. **105**(9), 097401 (2010).

12. J. A. Armstrong, N. Bloembergen, J. Ducuing, and P. S. Pershan, "Interactions between Light Waves in a Nonlinear Dielectric," Phys. Rev. **127**(6), 1918–1939 (1962).
13. S.-Y. Hong, J. I. Dadap, N. Petrone, P.-C. Yeh, J. Hone, and R. M. Osgood, "Optical Third-Harmonic Generation in Graphene," Phys. Rev. X **3**(2), 021014 (2013).
14. N. Kumar, J. Kumar, C. Gerstenkorn, R. Wang, H.-Y. Chiu, A. L. Smirl, and H. Zhao, "Third harmonic generation in graphene and few-layer graphite films," Phys. Rev. B **87**(12), 121406 (2013).
15. B. Zhao and Z. M. Zhang, "Strong Plasmonic Coupling between Graphene Ribbon Array and Metal Gratings," ACS Photonics **2**(11), 1611–1618 (2015).
16. P. B. Johnson and R. W. Christy, "Optical Constants of the Noble Metals," Phys. Rev. B **6**(12), 4370–4379 (1972).
17. M. Walther, D. G. Cooke, C. Sherstan, M. Hajar, M. R. Freeman, and F. A. Hegmann, "Terahertz conductivity of thin gold films at the metal-insulator percolation transition," Phys. Rev. B **76**(12), 125408 (2007).
18. A. Ishikawa, T. Tanaka, and S. Kawata, "Negative Magnetic Permeability in the Visible Light Region," Phys. Rev. Lett. **95**(23), 237401 (2005).
19. E. Poutrina, C. Ciracì, D. J. Gauthier, and D. R. Smith, "Enhancing four-wave-mixing processes by nanowire arrays coupled to a gold film," Opt. Express **20**(10), 11005 (2012).
20. G. W. Hanson, "Dyadic Green's Functions for an Anisotropic, Non-Local Model of Biased Graphene," IEEE Trans. Antennas Propag. **56**(3), 747–757 (2008).
21. D. Chatzidimitriou, A. Pitilakis, and E. E. Kriezis, "Rigorous calculation of nonlinear parameters in graphene-comprising waveguides," J. Appl. Phys. **118**(2), 023105 (2015).
22. Z. Li, K. Yao, F. Xia, S. Shen, J. Tian, and Y. Liu, "Graphene Plasmonic Metasurfaces to Steer Infrared Light," Sci. Rep. **5**, 12423 (2015).
23. I. Al-Naib, J. E. Sipe, and M. M. Dignam, "High harmonic generation in undoped graphene: Interplay of inter- and intraband dynamics," Phys. Rev. B **90**(24), 245423 (2014).
24. X. Li, C. W. Magnuson, A. Venugopal, R. M. Tromp, J. B. Hannon, E. M. Vogel, L. Colombo, and R. S. Ruoff, "Large-Area Graphene Single Crystals Grown by Low-Pressure Chemical Vapor Deposition of Methane on Copper," J. Am. Chem. Soc. **133**(9), 2816–2819 (2011).
25. M. Kahl, E. Voges, S. Kostrewa, C. Viets, and W. Hill, "Periodically structured metallic substrates for SERS," Sensors Actuators B Chem. **51**(1–3), 285–291 (1998).
26. J. L. Cheng, N. Vermeulen, and J. E. Sipe, "Third order optical nonlinearity of graphene," New J. Phys. **16**(5), 053014 (2014).
27. J. Niu, M. Luo, and Q. H. Liu, "Enhancement of graphene's third-harmonic generation with localized surface plasmon resonance under optical/electro-optic Kerr effects," J. Opt. Soc. Am. B **33**(4), 615 (2016).
28. B. Jin, T. Guo, and C. Argyropoulos, "Enhanced third harmonic generation with graphene metasurfaces," J. Opt. **19**(9), 094005 (2017).
29. S. Thongrattanasiri, F. H. L. Koppens, and F. J. García de Abajo, "Complete Optical Absorption in Periodically Patterned Graphene," Phys. Rev. Lett. **108**(4), 047401 (2012).
30. W. L. Barnes, W. A. Murray, J. Dintinger, E. Devaux, and T. W. Ebbesen, "Surface Plasmon Polaritons and Their Role in the Enhanced Transmission of Light through Periodic Arrays of Subwavelength Holes in a Metal Film," Phys. Rev. Lett. **92**(10), 107401 (2004).
31. P.-Y. Chen, M. Farhat, and H. Bağcı, "Graphene metascreen for designing compact infrared absorbers with enhanced bandwidth," Nanotechnology **26**(16), 164002 (2015).
32. H. Nasari and M. S. Abrishamian, "Nonlinear terahertz frequency conversion via graphene microribbon array," Nanotechnology **27**(30), 305202 (2016).
33. T. Guo, B. Jin, and C. Argyropoulos, "Hybrid Graphene-Plasmonic Gratings to Achieve Enhanced Nonlinear Effects at Terahertz Frequencies," Phys. Rev. Appl. **11**(2), 024050 (2019).
34. T. Guo, L. Zhu, P.-Y. Chen, and C. Argyropoulos, "Tunable terahertz amplification based on photoexcited active graphene hyperbolic metamaterials [Invited]," Opt. Mater. Express **8**(12), 3941 (2018).